contact: Alexander Kozhanov kozhanov@cnsi.ucsb.edu

# Micro-Structured Ferromagnetic Tubes for Spin Wave Excitation.

A. Kozhanov, D. Ouellette, M. Rodwell, S. J. Allen
*California Nanosystems Institute, University of California at Santa Barbara, Santa Barbara, CA, 93106*
D. W. Lee and S. X. Wang
*Department of Materials Science and Engineering, Sanford University, Stanford, CA, 94305*

September 24, 2010

Micron scale ferromagnetic tubes placed on the ends of ferromagnetic CoTaZr spin waveguides are explored in order to enhance the excitation of Backward Volume Magnetostatic Spin Waves. The tubes produce a closed magnetic circuit about the signal line of the coplanar waveguide and are, at the same time, magnetically contiguous with the spin waveguide. This results in a 10 fold increase in spin wave amplitude. However, the tube geometry distorts the magnetic field near the spin waveguide and relatively high biasing magnetic fields are required to establish well defined spin waves. Only the lowest (uniform) spin wave mode is excited.

(PACS: 76.50.+g)

Small scale magnetostatic spin wave devices are potentially important for various applications such as on-chip tunable filters for communication systems[1] and inductors[2,3] and spin wave logic devices[4,5,67]. Magnetostatic spin wave devices based on insulating ferrimagnetic materials like yttrium-iron garnet (YIG)[8] have been intensively explored and developed. However, significantly higher saturation magnetization of ferromagnetic metals like CoFe, NiFe and CoTaZr[9] as well as ease of film deposition and processing at the micron and nanometer scale has made these materials potentially more important. Strong shape anisotropy arises when the ferromagnetic metal films are patterned into wires or stripes. This results in finite frequency spin wave modes with little or no biasing magnetic fields; the dispersion of the backward volume magnetostatic spin waves (BVMSW's)[10] are determined by the profile dimensions. Ferromagnetic wires with magnetization, wave vector k and biasing magnetic field $H$ aligned along the wire have been explored theoretically[11,12] while elegant and powerful Brillouin light scattering experiments document thermally excited lateral standing wave patterns associated with the various spin wave modes.[13,14,15]

Most of the spin wave logic devices use coupling loops in form of shorted microwave transmission lines placed in vicinity of the ferromagnetic spin waveguide for the spin wave excitation[6,7,16]. Such excitation technique relies on the magnetic field produced by the current flowing through the signal line coupling loop and has been studied in detail[17,18]. In this geometry, the magnetic field produced by the coupling loop stays mainly outside of the spin waveguide because of the high magnetic permeability of the ferromagnetic material and results in inefficient spin wave excitation. This problem can be solved by closing magnetic circuit by wrapping the ferromagnetic material around the signal line. The closed magnetic circuit forms a ferromagnetic tube with magnetic flux concentrated within the walls which should lead to more effective spin wave excitation.

In this work we use the BVMSW modes in ferromagnetic spin waveguides to study the spin wave excitation and detection with the micron-scale ferromagnetic tubes lithographically formed at the ends of the spin waveguides. To estimate efficiency of such spin wave excitation method we reference our results to the data measured on the identical structures without tube-couplers.

Micron size ferromagnetic spin waveguides in form of stripes with ferromagnetic tubes at the ends and shorted coplanar waveguides (Fig.1) were fabricated in the following manner. 100 nm thick ferromagnetic $Co_{90}Ta_5Zr_5$ film was sputtered onto $Si/SiO_2$ wafers. A saturation magnetization of $M_s$=1.2 T and a coercive field $H_c$~3 Oe were measured on an unpatterned $Co_{90}Ta_5Zr_5$ film using a vibrating sample magnetometer. Deposited films were lithographically patterned into 2 and 4 μm wide stripes with lengths varying from 8 to 14 μm (distance between excitation and detection tubes vary from 4 to 12 μm). Patterned $Co_{90}Ta_5Zr_5$ film was covered with a 50 nm thick insulating $SiO_2$ layer. 100 nm thick coupling loops formed by shorting the ends of a pair of coplanar waveguides were positioned over the ends of $Co_{90}Ta_5Zr_5$ stripes. The structures were then covered with a 150 nm thick $SiO_2$ insulating layer and holes were etched to allow a subsequent top $Co_{90}Ta_5Zr_5$ layer to complete the magnetic circuit. The top $Co_{90}Ta_5Zr_5$ layer was sputtered on the resist covered structure and lifted-off. This process resulted in shorted coplanar waveguide signal lines running through a $Co_{90}Ta_5Zr_5$ tubes. Cross section obtained by etching part of fabricated structure with focused ion beam is shown in Figure.1b. Microwave currents flowing through the signal line of the shorted coplanar waveguide generate magnetic fields which are confined within the walls of the ferromagnetic tubes and used for the spin wave excitation. A schematic drawing of the spin wave excitation in the fabricated structure is shown in Figure 1c.

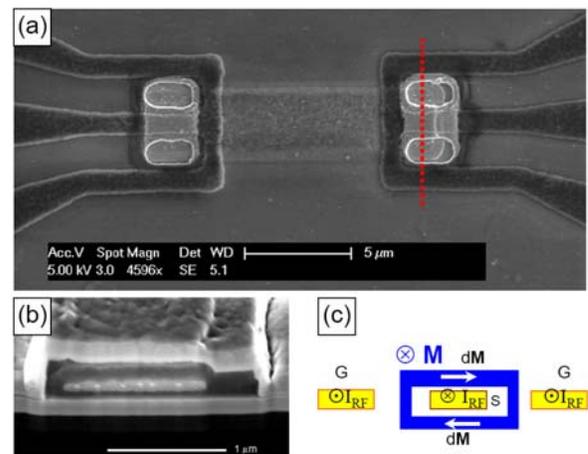

FIG.1. Fabricated structure: SEM micrograph: top view (a), SEM micrograph of the cross section along the dashed line in the top view SEM; cross section schematic (c)

S-parameters were measured at room temperature using Agilent 8720ES vector network analyzer. Only $S_{21}$, the ratio of high frequency voltage at terminals 2 to the input high frequency voltage at terminals 1, is analyzed in the following discussion. A GWV projected field magnet provided in-plane bias fields up to 2500 Oe. By comparing the S-parameters at disparate bias magnetic fields, the magnetic field independent instrument response can be effectively removed to expose the S - parameters of the magneto-static spin wave guide.

Typical results for the real and imaginary parts of $S_{21}$ measured on a structure with $(0.11 \times 3.75 \times 10)$ μm$^2$ $Co_{90}Ta_5Zr_5$ stripe at a finite magnetic field are shown in Figure 2a. Reproducible oscillations are seen on the real and imaginary parts of $S_{21}$ which shift to higher frequencies as the biasing magnetic field directed along the stripe length is increased. A $\pi/2$ phase offset of real and imaginary parts of $S_{21}$ and more than $20\pi$ changes in phase are recorded. The latter scale with the stripe length in the measured frequency range and are indicative of the magnetostatic spin wave traveling through the spin waveguide. The geometry used in this experiment favors excitation of BVMSW modes where the wave vector is parallel to the magnetization of the stripe. In contrast to the data obtained on the spin waveguides without tunnels at the ends[16] the structures measured in this work exhibit only one set of oscillations with continuous phase evolution. The closed magnetic circuit produces magnetization oscillations that favor

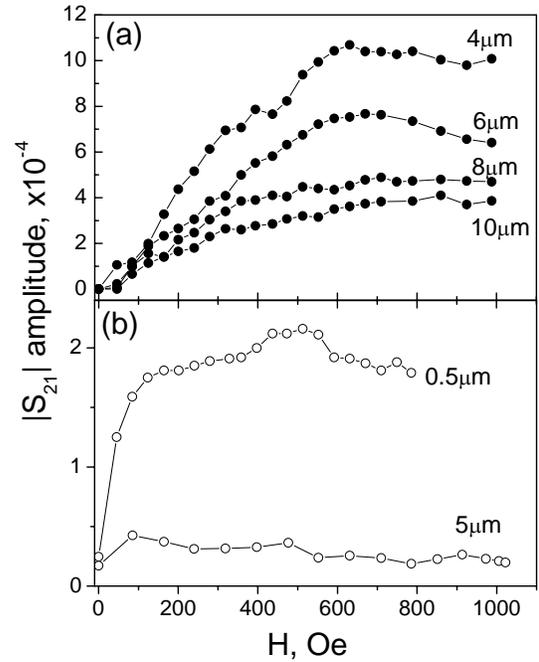

FIG.3. Magnetic field dependence of $|S_{21}|$ amplitude measured on spin wave-guide with (a) and without (b) ferromagnetic tubular spin wave couplers placed at the CoTaZr stripes. Numbers near curves show the distance between excitation and detection points.

excitation of the uniform mode and we assign the signals accordingly. (Fig.2).

Figure 3 shows magnetic field dependence of the transmitted signal amplitude for the structures with (Fig.3a) and without (Fig3b) tubular couplers measured at the same level of excitation current amplitude. Structures without ferromagnetic tubes exhibit weak magnetostatic coupling excitation at H=0. As the magnetic field is increased the spin wave amplitude goes through a step-like increase and saturates with H~100 Oe at a level of $|S_{21}|\sim 2\times 10^{-4}$. This indicates that even at H=0 some fraction of the magnetic moments in stripe is aligned along the stripe length which allows spin wave propagation even at zero magnetic field. Low spin wave amplitude at H=0 is caused by the magnetization distortions at the ends of the stripe which is used for the spin wave excitation and detection. This is confirmed by the micromagnetic simulations of the magnetization alignment of the stripe at H=0. A small magnetic field applied along the stripe length aligns the distorted magnetic moments increasing transmitted signal until the stripe is fully magnetized.

Placing tubular couplers at the end of the spin waveguides introduces strong magnetic disorder at the ends of the stripe that results in zero spin wave transmission at H=0 (Fig.3b). Spin waveguides with tubes at the ends require much higher biasing magnetic field to uniformly magnetize the spin waveguide than do the spin waveguides without tubes. Increasing the biasing magnetic field results in slow increase of the $|S_{21}|$ amplitude which reaches a saturated value only at H~700 Oe. This indicates that spin waveguides with tubular couplers at the ends require much higher biasing magnetic field for the spin wave propagation. The saturation level of the $|S_{21}|$ amplitude for the stripe with 5μm distance between exciting and detecting tubes is $|S_{21}|\sim 1\times 10^{-3}$ which is about one order of magnitude higher than measured in a structures of the same length but without tubular couplers. However, irrespective of the coupling structure, similar attenuation lengths of ~3μm were measured in similar spin waveguides, indicating that the improved transmission was due to coupling in and out. [16].

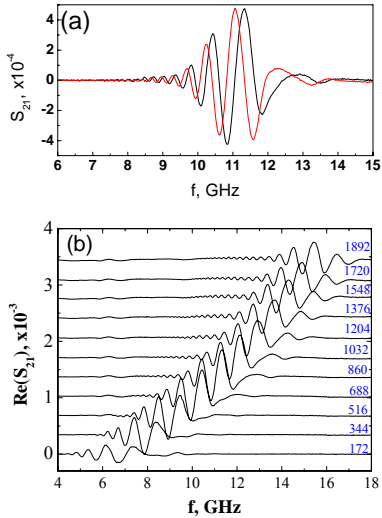

FIG.2. Frequency dependence of real and imaginary parts of $S_{21}$ measured on $(0.11\times 3.75\times 8)$μm$^3$ $Co_{90}Ta_5Zr_5$ stripes with 2μm long CoTaZr tubes at the ends, H=860 Oe. Frequency dependences of Re($S_{21}$) measured at different magnetic fields. Numbers near curves are magnetic field in Oe. (b).

In summary, we studied spin wave excitation by CoTaZr tubes placed at the ends of the spin waveguides. Ferromagnetic tubes at the ends of ferromagnetic wires form a closed magnetic circuit around the exciting and detecting coupling loop signal lines. This results in about one order of magnitude increase of the spin wave amplitude. However the shape anisotropy of the coupling tubes requires higher biasing magnetic fields (H>700Oe) for maximal spin wave transmission excitation. By making the tubes longer shape anisotropy should define the tube magnetization ground state to be magnetized along the tube axis. This should allow efficient spin wave excitation and detection at zero biasing magnetic fields.

This work is supported by Nano Electronics Research Corporation (NERC) via the Nanoelectronics Research Initiative (NRI), by Intel Corp. and UC Discovery at the Western Institute of Nanoelectronics (WIN) Center.